\documentclass[twocolumn,showpacs,preprintnumbers,showkeys,prd,nofootinbib,amsmath,amssymb]{revtex4}

\usepackage{graphicx}

\begin{document}

\def\half{\textstyle{1\over2}}
\def\third{\textstyle{1\over3}}
\def\quarter{\textstyle{1\over4}}

\newcommand{\be}{\begin{equation}}
\newcommand{\ee}{\end{equation}}
\newcommand{\bea}{\begin{eqnarray}}
\newcommand{\eea}{\end{eqnarray}}
\newcommand{\ud}{\mathrm{d}}

\title{Strong gravitational lensing by braneworld black holes}

\author{Richard Whisker}

\email{r.s.whisker@dur.ac.uk}

\affiliation{Institute for Particle Physics Phenomenology, University of
  Durham, South Road, DH1 3LE, UK.}

\date{\today}

\begin{abstract}
In this paper, we use the strong field limit approach to investigate the
gravitational lensing properties of braneworld black holes.  Applying this method to
the supermassive black hole at the centre of our galaxy, the lensing
observables for some candidate braneworld black hole metrics are compared with those for the
standard Schwarzschild case. It is found that braneworld black holes could
have significantly different observational signatures to the Schwarzschild
black hole.
\end{abstract}

\pacs{04.50.+h, 04.70.Bw, 95.30.Sf, 98.62.Sb}

\keywords{Black holes; Braneworlds; Gravitational lensing}

\preprint{IPPP/04/80}
\preprint{DCPT/04/160}

\maketitle

\section{Introduction} \label{sec:intro}
The braneworld paradigm provides an interesting framework within which to
explore the possibility that our Universe lives in a fundamentally higher
dimensional spacetime.  Unlike the Kaluza-Klein picture where the extra
dimensions must be compactified on a length scale $R \lesssim 10^{-17}$cm in
order to evade our detection, confinement of the standard model fields to a
3-brane, with only gravity propagating in the bulk, allows large, and even
infinite, extra dimensions. 

Recent work on braneworlds was instigated by the
proposals of Arkani-Hamed, Dimopoulos and Dvali (ADD)~\cite{ADD} and Randall
and Sundrum (RS)~\cite{RS} (see \cite{earlier} for early work, and
\cite{branerev} for braneworld reviews). The ADD model has $n$ flat, compact
extra dimensions of size $R$.  Due to the confinement of ordinary matter to a
brane, $R$ can be as large as $\sim \! 0.1$mm (the scale down to which Newton's
law has been experimentally tested), and the model provides a possible
resolution to the hierarchy problem if $n \ge 2$.  More interesting from the
general relativity viewpoint are the RS models proposed shortly afterwards.
They allowed the bulk geometry to be curved, and endowed the branes with a
tension.  Their first model consisted of two branes of equal but opposite
tension bounding a slice of Anti-de-Sitter space.  This model also gives a
possible resolution to the hierarchy problem, provided we live on the
negative tension brane. In their second model, RS considered a single,
positive tension brane in an \emph{infinite} bulk.  This model, which is
loosely motivated by string theory~\cite{HW},  has been
one of the most popular to explore, and is the one we will be using.  

Even though the extra dimension is infinite and gravity is inherently
five dimensional, RS showed that the Newtonian potential of a particle on the
brane was indeed the 4D $1/r$ potential.  This result was backed up by more
complete analyses which confirmed that the graviton propagator did indeed
have the correct tensor structure, and that the effect of the extra dimension
was to introduce a $1/r^3$ correction to the gravitational
potential~\cite{GT}:
\be
V(r) = \frac{G_N}{r} \left(1+\frac{2}{3}\frac{l^2}{r^2}\right).
\ee
An elegant description of non-perturbative gravity on the brane was provided
by Shiromizu, Maeda and Sasaki~\cite{SMS}.  Using a Gauss-Codazzi approach,
they projected the 5D Einstein equations onto the brane to obtain the
effective 4D field equations:      
\be \label{SMSeq} G_{\mu\nu} = \Lambda_4g_{\mu\nu} + 8\pi
G_NT_{\mu\nu} + \kappa^2 S_{\mu\nu} + {\cal E}_{\mu\nu}. \ee
Here, $\Lambda_4$ is a residual cosmological constant on the brane and
represents the mismatch between the brane tension and the negative bulk
cosmological constant.  $T_{\mu\nu}$ is the usual energy-momentum tensor of
matter on the brane, and $S_{\mu\nu}(T^2)$ consists of squares of
$T_{\mu\nu}$ and thus is a local, high energy correction term.  ${\cal
  E}_{\mu\nu}$ consists of the projection of the bulk Weyl tensor onto the
brane, and is non-local from the brane point of view.  It is important to
emphasise that since ${\cal E}_{\mu\nu}$ is not given in terms of data on the
brane, the system of equations (\ref{SMSeq}) is not closed, in general.  

The generalisation of the FRW Universe that follows from Eq.~(\ref{SMSeq}) has
been well explored~\cite{Cos}.  The $S_{\mu\nu}$ term contributes a high energy
correction term to the Friedmann equation, which is relevant only in the very
early Universe, and the Weyl term contributes a `dark radiation' term.
Hence, although all the implications may not have been calculated, braneworld
cosmology for the pure RS scenario is pretty well understood.  The situation
for braneworld black holes (BBHs) is somewhat more complicated however, and
there is no longer a simple solution~\cite{CHR,bbh}.  Black holes are fascinating
objects, and provide a potential testing ground for general relativity.  It
is therefore important to investigate braneworld generalisations of the
Schwarzschild solution, and the possible observational signatures that could
result. 

The theory of gravitational lensing has been mostly developed in the weak
field approximation, where it has been successful in explaining all
observations~\cite{lensrev}.  However, one of the most spectacular
consequences of the strong gravitational field surrounding a black hole is
the large bending of light that can result for a light ray passing through
this region.  The study of 
strong gravitational lensing was resurrected recently by Virbhadra and
Ellis~\cite{Virb}, who studied lensing by the galactic supermassive black
hole, in an asymptotically flat background.  Frittelli, Kling and
Newman~\cite{Frit} found an exact lens equation without reference to a
background metric and compared their results with those of Virbhadra and
Ellis.  In~\cite{Bozza1} Bozza \emph{et al}.\ first defined a strong field
limit and used it to investigate Schwarzschild black hole lensing
analytically.  This technique has been applied to Reissner-Nordstr\"om black
holes~\cite{Eiroa} and the GMGHS~\cite{gmghs} charged black hole of heterotic string
theory~\cite{Bhadra}, and was generalised to an arbitrary static, spherically
symmetric metric by Bozza~\cite{Bozza}.  In this paper we utilise this method
to investigate the gravitational lensing properties of a couple of candidate
BBH metrics.  

Similar studies have been performed for a BBH with the induced geometry of
the 5D Schwarzschild solution: $g_{tt} = g_{rr}^{-1} = 1-r_h^2/r^2$.  Both weak
field lensing~\cite{Majumdar} and strong field lensing~\cite{Eiroa2} for this
geometry have been studied.  However, this metric is only appropriate for black
holes with a horizon size smaller than the AdS length scale of the extra
dimension:
$r_h<l \lesssim 0.1$mm. Hence, this metric is not appropriate for
investigating the phenomenology of massive astrophysical black holes.

\section{Braneworld black holes} \label{sec:BBH}
The general static, spherically symmetric metric on the brane can
be written as: \be \label{metric} \ud s^2 = g_{\mu\nu} \ud x^{\mu}
\ud x^{\nu} = A^2(r) \ud t^2 - B^2(r)\ud r^2 - C^2 (r) \ud
\Omega_{I \!\! I}^2. \ee Clearly, this is not in the simplest
gauge, as we can still choose our radial coordinate, $r$, quite
arbitrarily.  The vacuum brane field equations following from
Eq.~(\ref{SMSeq}) (with $\Lambda_4$ set equal to zero) are \be
\label{vac} G_{\mu\nu} = {\cal E}_{\mu\nu}. \ee The solution of
these equations requires the input of ${\cal E}_{\mu\nu}$ from the
full 5D solution.  In the absence of such a solution, an
assumption about ${\cal E}_{\mu\nu}$ or $g_{\mu\nu}$ must be made
in order to close the system of equations.

Several special solutions, making various assumptions about
$g_{\mu\nu}$, have been presented in the literature.  The first
attempt at a BBH solution was the so-called black string solution
of Chamblin \emph{et al.}~\cite{CHR}, which consists simply of the 4D
Schwarzschild solution `stacked' into the extra dimension.
Unfortunately, this solution has a singular AdS horizon and is
unstable to classical perturbations \cite{GL}. The assumption $A^2 = 1/B^2$
leads to the tidal Reissner-Nordstr\"om solution of Dadhich \emph{et al.}
~\cite{RN}: \bea \label{rneq} \ud s^2 \! &=& \! \left( 1-{2GM \over r}+{Q \over
r^2} \right) \ud t^2 - \left(
  1-{2GM \over r}+{Q \over r^2} \right)^{-1} \! \! \ud r^2 \nonumber \\
&& {} - r^2 \ud \Omega_{I \!\! I}^2.
\eea Unlike the standard Reissner-Nordstr\"om solution, the `tidal
charge' parameter $Q$ can take both positive \emph{and} negative
values.  Indeed, negative $Q$ is the more natural since
intuitively we would expect the tidal charge to \emph{strengthen}
the gravitational field, as it arises from the source mass $M$ on
the brane (see~\cite{RN} for further discussion). This metric has the correct
5D ($\sim \!  1/r^2$) short
distance behaviour and so could be a good approximation in the
strong field regime for small black holes.  Solutions have also
been found which assume a given form for the time
or radial part of the metric \cite{CAS,GM}.  Visser and Wiltshire
\cite{VW} presented a more general method which generated an exact solution
for a given radial metric form.

In all the above cases the radial gauge $C=r$ was chosen (although
\cite{VW} did comment on how to use their method when $C(r)$ was
not monotonic).  However, there are good reasons to believe that
the area ${\cal A}$ of the 2-spheres might not be monotonic. The
second derivative of the area radius $C$ (i.e.~the radial function
defined by $\sqrt{{\cal A}/4 \pi}$) is given by \be {C'' \over C}
= -{B^2 \over 2}(G^t_t - G^r_r) + {C' \over C}\left({B' \over
    B}+ {A' \over A} \right).
\ee Hence for the area function to be guaranteed to be monotonic we must have
$G_t^t-G^r_r \geq 0$, which is equivalent to the dominant energy
condition.  While this is generally satisfied in standard Einstein
gravity, it need not be in the case of extra dimensions, and so it
is important for the exploration of BBHs that we do not make the
restrictive ansatz $C=r$. (For a discussion of non-monotonic
radial functions in the context of braneworld \emph{wormhole}
solutions, see \cite{Bronn}.)

An alternative to making guesses for the metric functions
$g_{\mu\nu}$ is instead to make an assumption about the Weyl term
${\cal
  E}_{\mu\nu}$. Although ${\cal E}_{\mu\nu}$  is a complete unknown from the
brane point of view, the symmetry of the problem allows it to be
decomposed as \cite{Maar} \be {\cal E}_{\mu\nu} = {\cal U}\left(u_\mu u_\nu -
\third h_{\mu\nu} \right) + \Pi \left(r_\mu r_\nu + \third
h_{\mu\nu} \right), \ee where $u^\mu$ is a unit time vector and
$r^\mu$ is a unit radial vector.
Recently, we proposed a pragmatic approach to BBHs, in which an
equation of state for the Weyl term is assumed \cite{us} (see also \cite{HM}): \be \Pi = {\gamma
-1 \over 2}\, {\cal U}. \ee Of course, a priori there is no reason
to suppose that the Weyl term should obey an equation of state.
However, it is quite possible that it might have certain
asymptotic equations of state which may be useful as near-horizon
or long-range approximations to the (as yet unknown) exact
solution.  Using a dynamical systems approach, the system of
equations (\ref{vac}) was solved, and the behaviour of the
solutions classified according to the equation of state parameter
$\gamma$. It was found that asymptotically flat solutions require
an equation of state $\gamma < 0$, and for $|\gamma| > 3$ the BBH
solutions have a singular horizon, and are allowed both with and
without turning points in the area function.

Using holography considerations, it was argued that we might
expect equations of state with large $\gamma$ to be relevant near
the horizon.  Taking this reasoning to its extreme, we proposed as
a `working' metric for the near-horizon geometry the ${\cal U} =
0$ (i.e.~$\gamma = \pm \infty$) solution: \be \label{U=0} \ud
s^2 = \frac{(r-r_h)^2}{(r+r_t)^2}\, \ud t^2 -
\frac{(r+r_t)^4}{r^4}\, \ud r^2 - \frac{(r+r_t)^4}{r^2}\, \ud
\Omega^2_{I\!\!I}, \ee which has a turning point in the area
function at $r=r_t$, and the horizon at $r=r_h$ is singular
(except for the special case $r_h=r_t$, which is just the standard
Schwarzschild solution in isotropic coordinates).  This metric has
appeared in the area gauge as \cite{CAS} \bea \label{areagauge} \ud s^2 &=&
\left( (1+\epsilon)\sqrt{1 - {2GM \over R}} - \epsilon \right)^2 \!
\ud t^2 \nonumber \\ && {} - \left(1 - {2GM \over R} \right)^{-1} \! \ud R^2
 - R^2 \ud \Omega_{I\!\!I}^2, \eea where $R = (r+r_t)^2/r$, $GM=2r_t$ and
$GM\epsilon = r_h-r_t$. For $\epsilon
> 0$ this gauge is valid in the whole horizon exterior, however for $\epsilon
< 0$ the turning point $r_t$ is outside the horizon $r_h$ and so
the area gauge is inappropriate. 

Although the choice of metric (\ref{U=0}) is somewhat arbitrary,
we believe that the horizon is likely to be singular and that a
turning point in the area function is also likely, and this metric
exhibits both these features.  It has the added advantage of being
analytic, and so seems a good choice for exploring these radical
differences to the standard Schwarzschild geometry. In this paper
we investigate the gravitational lensing properties of the metrics
(\ref{rneq})\footnote{In the context of the equation
of state, (\ref{rneq}) has $\gamma = -3$.} and (\ref{U=0}), to see how braneworld
effects might manifest themselves in observations of black holes.

It is important to emphasise that we envisage these only as
possible near-horizon asymptotes of a more general metric, which
has yet to be found. Neither metric satisfies the long distance
$1/r^3$ correction to the gravitational potential, and both would
be constrained in the weak field by the PPN observations.

\section{Gravitational lensing} \label{sec:gravlens}
\begin{figure}
\begin{center}
\begin{picture}(0,0)
\includegraphics[scale=0.8]{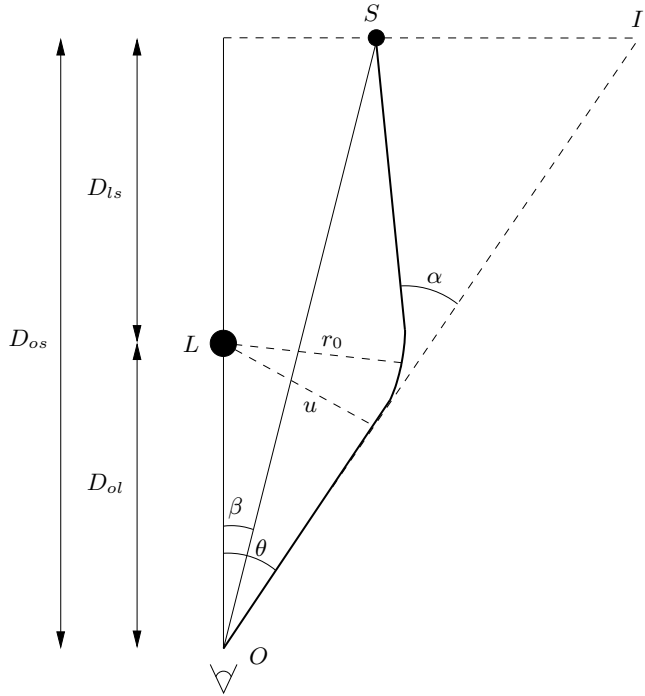}
\end{picture}
\setlength{\unitlength}{0.8cm}
\begin{picture}(10,11)(0.15,0)
\put(3.1,0.5){$O$}
\put(2,5.7){$L$}
\put(5,11.2){$S$}
\put(9.45,11.1){$I$}
\put(2.75,3){$\beta$}
\put(3.2,2.3){$\theta$}
\put(6.05,6.85){$\alpha$}
\put(4,4.7){$u$}
\put(4.3,5.8){$r_0$}
\put(0.4,3.4){$D_{ol}$}
\put(0.4,8.3){$D_{ls}$}
\put(-0.9,5.8){$D_{os}$}
\end{picture}
\end{center}
\caption{Gravitational lensing diagram.\label{fig:lens}}
\end{figure}
The lensing setup is shown in Fig.~\ref{fig:lens}.
Light emitted by the source $S$ is deflected by the lens $L$ and
reaches the observer $O$ at an angle $\theta$ to the optic axis
$OL$, instead of $\beta$.  The spacetime, described by the metric
(\ref{metric}) centred on $L$, is asymptotically flat, and both
observer and source are located in the flat region. By simple
trigonometry, the lens equation can be written down: \be
\label{lenseq} \tan \beta = \tan \theta - {D_{ls} \over
D_{os}}\left[ \tan \theta + \tan (\alpha - \theta) \right]. \ee
From the null geodesic equations it is straightforward to show
that the angular deflection of light as a function of radial distance from
the lens is \be {\ud \phi \over \ud r} = {B \over C \sqrt{{C^2 \over
u^2A^2} -1}}. \ee By conservation of angular momentum, the impact
parameter $u$ is given by \be u = {C_0 \over A_0}, \ee where the
subscript $0$ indicates that the function is evaluated at the
closest approach distance $r_0$.  Hence, the deflection angle is
given by \bea \alpha(r_0) &=& I(r_0) \, - \, \pi \nonumber \\
&=&
\int_{r_0}^{\infty} {2B \over C} \left({C^2 \over C_0^2}{A_0^2 \over A^2}
  -1\right)^{\!-{1 \over 2}} \! \ud r \, - \, \pi. \label{def} \eea Equations
(\ref{lenseq})
and (\ref{def}) are the basic equations of gravitational lensing. In
principle, the deflection angle $\alpha$ for a given metric can be
calculated from Eq.~(\ref{def}). This can then be plugged into the
lens equation (\ref{lenseq}), and the image position $\theta$ for
a given source position $\beta$ can be found.

The theory of gravitational lensing has been developed mostly in
the weak field limit, where several simplifying assumptions can be
made.  The angles in Eq.~(\ref{lenseq}) are taken to be small, so that
$\tan x$ can be replaced by $x$ for $x = \beta, \theta, \alpha$,
and the integrand in Eq.~(\ref{def}) is expanded to first order in the
gravitational potential.  For the Schwarzschild geometry, and
setting $\beta = 0$, this leads to the well-known result: \be
\theta_E = \sqrt{{4GM \over c^2}{D_{ls} \over D_{os}D_{ol}}} \, ,
\ee where $\theta_E$ is the Einstein radius.  In this formulation,
general relativity has been successful in explaining all
observations (see \cite{lensrev} for detailed reviews).  However, it is
important that gravitational lensing is not conceived of as a
purely weak field phenomenon.  Indeed, gravitational lensing in
strong fields is one of the most promising tools for testing
general relativity in its full, non-linear form.

\subsection{Strong field limit} \label{sec:sfl}
\begin{figure}
\begin{center}
\resizebox{\hsize}{!}{\includegraphics{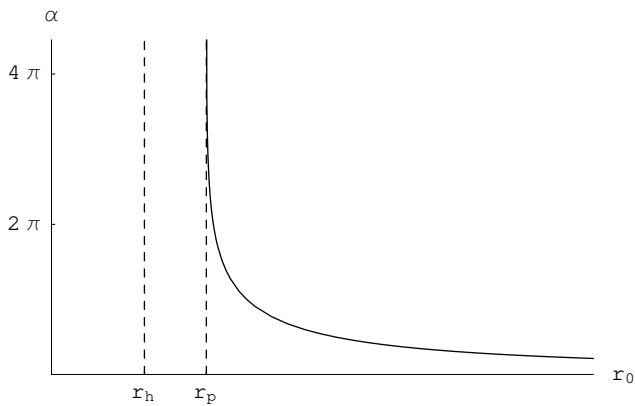}}
\end{center}
\caption{General behaviour of the deflection angle as a
function of $r_0$. As $r_0$ decreases, $\alpha$ increases, and each time it
reaches a multiple of $2\pi$ the photon completes a loop around the black
hole.\label{fig:deflec}}
\end{figure}

As the impact parameter $u$ of a light ray decreases, the
deflection angle $\alpha$ increases as shown in
Fig.~\ref{fig:deflec}.
At some point, the deflection angle exceeds $2 \pi$ and the photon
performs a complete loop around the black hole before emerging.
The images thus formed are termed `relativistic images' and a
theoretically infinite number of such images are formed on either
side of the lens, corresponding to successive winding numbers
around the black hole. The photon sphere is the radius $r_p$ at
which a photon can unstably orbit the black hole, and is defined
as the largest solution to the equation \be \label{ps} {A'(r)
\over A(r)} = {C'(r) \over C(r)}. \ee As $r_0$ approaches $r_p$,
with corresponding impact parameter \be u_p = {C_p \over A_p},
\label{up} \ee the deflection angle diverges and for $r_0<r_p$ the
photon is captured by the black hole.

Bozza \cite{Bozza} has shown that this divergence is logarithmic
for all spherically symmetric black hole metrics of the form
(\ref{metric}). Hence the deflection angle can be expanded close
to the divergence in the form \be \alpha (r_0) = -a \ln \left(
{r_0 \over r_p} -1 \right) + b + O(r_0-r_p), \ee or in terms of
the angular position of the image, $\theta = u/D_{ol}$, \be \label{SFL}
\alpha (\theta) = -\bar{a} \ln
\left( {\theta D_{ol} \over u_p} -1 \right) + \bar{b} + O(u-u_p), \ee where
the \emph{strong field limit} (SFL) coefficients $\bar{a}$ and
$\bar{b}$ depend on the metric functions evaluated at $r_p$.  This
formula allows a simple, analytic description of the relativistic
images and their properties, rather than having to use the exact
deflection angle calculated numerically from Eq.~(\ref{def}).

Equation (\ref{SFL}) can be derived from Eq.~(\ref{def}) by splitting
the integral into a divergent and non-divergent piece, and
performing some expansions.  Defining the new variable \be z =
{A^2 - A^2_0 \over 1-A^2_0}, \ee the integral in equation
(\ref{def}) becomes \be \label{def2} I(r_0) = \int_0^1
R(z,r_0)f(z,r_0) \ud z, \ee where \bea
R(z,r_0) &=& {B C_0 \over C^2 A'}(1-A^2_0), \\
f(z,r_0) &=& \left(A^2_0 - \left[(1-A^2_0)z + A^2_0\right]{C_0^2
\over C^2} \right)^{-{1 \over 2}}. \eea The function $R(z,r_0)$ is
regular for all values of $z$ and $r_0$, but $f(z,r_0)$ diverges
for $z \to 0$. Expanding the argument of the square root in
$f(z,r_0)$ to second order in $z$: \be f(z,r_0) \sim f_0(z,r_0) =
\left( m(r_0)z + n(r_0)z^2 \right)^{-1/2}, \ee \bea m(r_0) &=&
{A_0(1-A_0^2) \over A'_0}\left({C'_0 \over C_0}- {A'_0 \over
A_0}\right),\\ n(r_0) &=& {(1-A_0^2)^2 \over 4 A_0 A'_0
C_0}\bigg[3C'_0\left(1-{A_0 C'_0 \over A'_0 C_0}\right) \nonumber 
\\ && {} + {A_0 \over A'_0}\left(C''_0-{C'_0 A''_0 \over A'_0}\right)\bigg] ,
\eea it is clear why the deflection angle diverges logarithmically
for $r_0=r_p:$ with $r_p$ given by Eq.~(\ref{ps}), $m(r_p)$ vanishes.
Hence for $r_0=r_p$, $f_0 \propto 1/z$ and the integral
(\ref{def2}) diverges logarithmically.

Proceeding to split the integral (\ref{def2}) into a divergent and
a regular piece, and performing further expansions (see
\cite{Bozza} for the detailed derivation\footnote{Note: The
expressions in \cite{Bozza} look slightly different to ours since
we define the metric as
$g_{\mu\nu}={\mathrm{diag}}(A^2,-B^2,-C^2)$ as opposed to
$g_{\mu\nu}={\mathrm{diag}}(A,-B,-C)$ in \cite{Bozza}.}), the SFL
coefficients are obtained:
\bea \bar{a} &=& {R(0,r_p) \over 2 \sqrt{n_p}}, \label{abar} \\
\bar{b} &=& \bar{a} \ln {2n_p \over A^2_p} + b_R - \pi,
\label{bbar} \eea where \be n_p = n(r_p) = {(1-A_p^2)^2 \over 4
C_p A'^3_p}(C''_p A'_p - C'_p A''_p) \ee and \be b_R = \int_0^1
\left[R(z,r_p)f(z,r_p) - R(0,r_p)f_0(z,r_p)\right] \ud z, \ee is
the integral $I(r_p)$ with the divergence subtracted.

\section{Braneworld black hole lensing} \label{sec:bbh}
In this section we apply the method outlined in the previous
section to calculate the deflection angle in the strong field
limit for the candidate BBH metrics (\ref{rneq}) and (\ref{U=0})
discussed in Section~\ref{sec:BBH}.

\subsection{$U=0$ metric}
The $U=0$ metric (\ref{U=0}) has a number of key differences to the
standard Schwarzschild geometry; the horizon is singular, and the
area function can have a turning point that lies either inside or
outside the horizon. Another difference, due to the fact that
$g_{tt} \ne g_{rr}^{-1}$, is that the ADM mass and gravitational
mass (defined by $g_{tt}$) are no longer the same.

\begin{figure}
\begin{center}
\resizebox{\hsize}{!}{\includegraphics{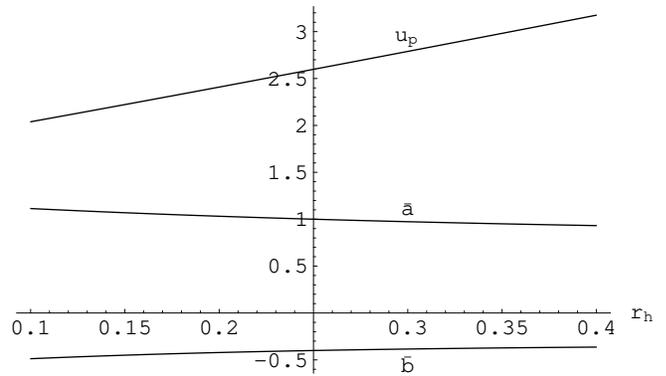}}
\end{center}
\caption{SFL coefficients for the $U=0$ metric (\ref{U=02}), as
 functions of $r_h$. The standard Schwarzschild case is given by $r_h = 1/4$.
\label{fig:U0}}
\end{figure}

Normalising the distances to $4 r_t$ (which corresponds to a
distance $2GM$ where $M$ is the ADM mass -- see
Eq.~(\ref{areagauge})), the metric functions are \bea A^2(r) &=&
\frac{(r-r_h)^2}{(r+1/4)^2}, \nonumber \\ B^2(r) &=&
\frac{(r+1/4)^4}{r^4}, \label{U=02}\\
C^2(r) &=& \frac{(r+1/4)^4}{r^2}. \nonumber \eea The radius of the
photon sphere is given by:
\be r_p ={1 \over 4}\left(1+4r_h + \sqrt{1+4r_h+16r_h^2}\right).\ee
 The SFL coefficients $\bar{a},\bar{b}$ and $u_p$, calculated from
Eqs.~(\ref{abar}),(\ref{bbar}) and (\ref{up}), are shown in
Fig.~\ref{fig:U0}.
It can be seen that the biggest deviation from standard
Schwarzschild lensing is for the minimum impact parameter $u_p$.
This is because as the horizon is shifted inwards/outwards
relative to the Schwarzschild case, the photon sphere is
pulled/pushed along with it. 

We can check the accuracy of the SFL approximation by comparing the exact
deflection angle $\alpha_{\mathrm{exact}}$ calculated from Eq.~(\ref{def}) with
the SFL $\alpha_{\mathrm{SFL}}$ from Eq.~(\ref{SFL}).  The outermost relativistic
image appears where $\alpha \simeq 2 \pi$, which occurs for an impact
parameter $u_1 = u_p + x$, where $x \sim 0.003$ depends on $r_h$. The discrepancy between
$\alpha_{\mathrm{exact}}(u_1)$ and  $\alpha_{\mathrm{SFL}}(u_1)$  is less
than $0.13 \%$ for all values of $r_h$ we consider.  Hence, the SFL of the
deflection angle is very accurate and can be reliably used to obtain accurate
results for the properties of the relativistic images.

\subsection{Tidal Reissner-Nordstr\"om metric}
\begin{figure}
\begin{center}
\resizebox{\hsize}{!}{\includegraphics{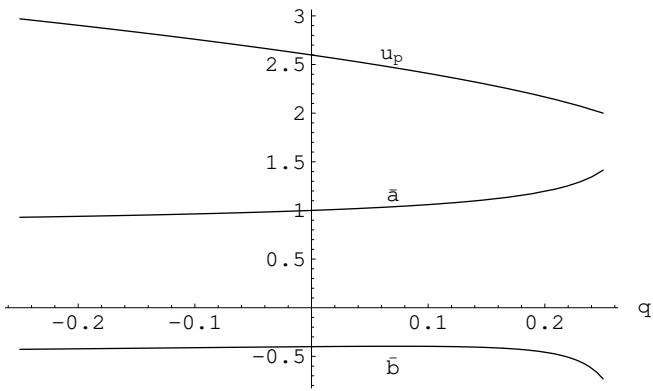}}
\end{center}
\caption{SFL coefficients for the Tidal Reissner-Nordstr\"om
  metric (\ref{rneq2}), as functions of $q$. The standard Schwarzschild case
  given by $q=0$.\label{fig:rn}}
\end{figure}

The tidal RN metric (\ref{rneq}) has the same properties as the standard RN
geometry for $q>0$: there are two horizons, both of which lie within the
Schwarzschild horizon, and the singularity at $r=0$ is timelike. However, we
can now have $q<0$ in which case there is just one horizon, lying outside
Schwarzschild, and the central singularity is spacelike, as in the
Schwarzschild case.  Normalising the distances to $2GM$, with $q =
Q/(2GM)^2$, the  metric (\ref{rneq}) becomes
\bea A^2(r) &=&
1-{1 \over r} + {q \over r^2}, \nonumber \\ B^2(r) &=&
\left(1-{1 \over r} + {q \over r^2}\right)^{-1}, \label{rneq2}\\
C^2(r) &=& r^2. \nonumber \eea
The radius of the photon sphere is given by:
\be r_p = {1 \over 4}\left(3+\sqrt{9-32q}\right). \ee
 The SFL coefficients are shown in Fig.~\ref{fig:rn}.
These results reproduce those of Eiroa \emph{et al.}~\cite{Eiroa} for $q>0$ (see
also~\cite{Bozza}), but we have extended the results to negative
$q$. We emphasise that there is \emph{no} electric charge present for the tidal RN
solution -- $q$ is a tidal charge parameter arising from the bulk Weyl
tensor. 

Again, comparing $\alpha_{\mathrm{exact}}$ with $\alpha_{\mathrm{SFL}}$
for an impact parameter corresponding to the outermost image, it is found
that the discrepancy is less than $0.5 \%$ for the values of $q$ considered here.

\section{Observables} \label{sec:obs} In Section \ref{sec:sfl} it was shown
how to calculate the deflection angle in the strong field limit and in
Section \ref{sec:bbh} this method was applied to the candidate BBH metrics.
In this section we put the SFL of the deflection angle into the lens equation
to obtain analytic formulae for the properties of the relativistic images in
terms of the SFL coefficients $\bar{a},\bar{b}$ and $u_p$.

As expected, the relativistic images formed by light rays winding
around the black hole are greatly de-magnified compared to the
standard weak field images, and are most prominent when the
source, lens and observer are highly aligned~\cite{Virb}. Hence,
we restrict our attention to the case where $\beta$ and $\theta$
are small (see~\cite{Manc} for the general case where this
assumption is relaxed). Although we can not assume $\alpha$ is
small, if a light ray is going to reach the observer after winding
around the black hole, $\alpha$ must be very close to a multiple
of $2\pi$. Writing $\alpha = 2n\pi +\Delta \alpha_n,\, n \in
\mathbb{Z},$ the lens equation (\ref{lenseq}) becomes \be \beta =
\theta - {D_{ls} \over D_{os}}\Delta \alpha_n . \ee

Firstly, we have to find the values $\theta_n^0$ such that
$\alpha(\theta_n^0) = 2n\pi$.  With $\alpha$ given by Eq.~(\ref{SFL})
we find \be \theta_n^0 = {u_p \over D_{ol}}(1+e_n), \ee where \be
e_n = e^{(\bar{b}-2n\pi)/\bar{a}}. \ee Thus the position of the
$n^{\mathrm{th}}$ relativistic image can be approximated
by~\cite{Bozza} \be \label{pos2} \theta_n = \theta_n^0 + {u_p e_n
  (\beta - \theta_n^0)D_{os} \over \bar{a} D_{ls}D_{ol}}, \ee
where the correction to $\theta_n^0$ is much smaller than
$\theta_n^0$. Approximating the position of the images by
$\theta_n^0$, the magnification of the $n^{\mathrm{th}}$
relativistic image is given by \be \label{mag2} \mu_n = {1 \over
(\beta/\theta)\partial \beta / \partial
  \theta}\Big|_{\theta_n^0} \simeq {u_p^2 e_n(1+e_n)D_{os} \over \bar{a} \beta
  D_{ol}^2 D_{ls}}. \ee
Equations (\ref{pos2}) and (\ref{mag2}) relate the position and
magnification of the relativistic images to the SFL coefficients.
We now focus on the simplest situation, where only the outermost
image $\theta_1$ is resolved as a single image, with the remaining
images packed together at $\theta_{\infty}=u_p/D_{ol}.$  Therefore
we define the observables \bea
s &=& \theta_1 - \theta_{\infty},\\
f &=& {\mu_1 \over \sum_{n=2}^{\infty} \mu_n}, \eea which are
respectively the separation between the outermost image and the
others, and the flux ratio between the outermost image and all the
others. It is found that these simplify to~\cite{Bozza}: \bea
s &=& \theta_{\infty} e^{(\bar{b}-2\pi)/\bar{a}},\\
f &=& e^{2\pi / \bar{a}}. \eea These equations are easily inverted
to give $\bar{a},\bar{b}$ and so if an observation were able to
measure $s, f$ and $\theta_{\infty}$ the SFL coefficients could be
determined and the nature of the lensing black hole identified.

\subsection{An Example: The galactic supermassive black hole}
It is believed that the centre of our galaxy harbours a
black hole of mass \mbox{$M  =  2.8 \! \times \!
10^6  M_{\odot}$~\cite{rich}}. Taking $D_{ol} = 8.5\,$kpc,
Virbhadra and Ellis~\cite{Virb} studied the lensing of a
background source by this black hole and found that the
relativistic images are formed at about $17 \, \mu$ arc sec.\ from
the optic axis.

In Table \ref{tab:table}
\begin{table*}
\begin{tabular}{|l|c|c|c|c|c|c|c|c|c|}

\hline 
{} & Schwarzschild & \multicolumn{4}{|c|}{$U=0$} & \multicolumn{4}{|c|}{Tidal
  RN}\\
\cline{3-10}
{} & {} & $r_h=0.1$ & $r_h=0.2$ & $r_h=0.3$ & $r_h=0.4$ & $q=-0.2$ 
& $q=-0.1$ & $q=0.1$ & $q=0.2$ \\
\hline
$\theta_{\infty}$ ($\mu$ arc sec.) & 16.87 & 13.24 & 15.65 & 18.11 & 20.62 & 18.85 & 17.92 &
15.64 & 14.07\\
$s$ ($\mu$ arc sec.) & 0.0211 & 0.0303 & 0.0235 & 0.0192 & 0.0164 & 0.0150 & 0.0173 & 0.0286 &
0.0502\\
$f_m$ (mags.) & 6.82 & 6.13 & 6.61 & 7.01 & 7.32 & 7.26 & 7.08 & 6.44 & 5.70\\
\hline 
\end{tabular}
 \caption{Estimates for the lensing observables for the central
 black hole of our galaxy.  $\theta_{\infty}$ and $s$ are defined in Section
 \ref{sec:obs}, and $f_m = 2.5 \log f$ is $f$ converted to magnitudes. \label{tab:table}}
\end{table*}
we estimate the observables $\theta_{\infty}, s, f$ defined in the
previous section for the $U=0$ and tidal RN BBH metrics, as well
as the standard Schwarzschild metric. It is clear that the easiest
observable to resolve is $\theta_{\infty}$, since a microarcsecond
resolution is in principle attainable by VLBI
projects such as MAXIM~\cite{max} and ARISE~\cite{arise}. However, the
disturbances inherent in such
observations would make the identification of the faint
relativistic images very difficult, as discussed in~\cite{Virb}.

If a measurement of $\theta_{\infty}$ was made, it would be
immediately capable of distinguishing between Schwarzschild and
other types of geometry.  However, to determine all the SFL
coefficients and thus unambiguously identify the nature of the
lensing black hole, it is necessary to also measure $s$ and $f$.
This would require the resolution of two extremely faint images
separated by $\sim 0.02 \, \mu$ arc sec. Such an observation in a
realistic astrophysical environment is certainly not feasible in
the near future, although if such an observation were ever
possible, it would provide an excellent test of gravity in a
strong field.

\section{Conclusions}
Gravitational lensing in strong fields provides a potentially
powerful tool for testing general relativity, and the strong field
limit provides a useful framework for comparing lensing by
different black hole metrics.  Of the possible alternatives to
standard GR, braneworld gravity is a very interesting model to
explore given the current interest in theories with extra
dimensions.  

In this paper we have investigated strong field lensing by potential
near-horizon BBH metrics.  Although the correct BBH metric is unknown and
much theoretical work remains to be done, this study is a useful first step
to explore the possible effects that braneworlds could have on the spacetime
surrounding a black hole. 

Table \ref{tab:table} clearly shows that BBHs could
have significantly different observational signatures than the
standard Schwarzschild black hole. Although the resolutions
required for these observations are beyond reach of current observational facilities,
this encourages the investigation of more realistically observable
situations.  An interesting possibility in this direction is the study of the
accretion discs surrounding black holes.  

The observed disc emission depends
on several factors that could get modified by braneworld effects.  A key
factor is the radius of the innermost stable circular orbit, since emitting
material at this radius sets the maximum temperature for the disc emission~\cite{acc1}.
Just as the radius of the photon sphere is shifted inwards or outwards
relative to the Schwarzschild case for a BBH, so too is the radius of the
innermost stable orbit for matter~\cite{us}.  In addition, the observed disc
emission, and in particular the iron fluorescence line profile, is affected
by relativistic effects~\cite{acc2} (doppler shift, gravitational redshift), which would
be modified if the metric in the emitting region was not that of standard
general relativity, but was a modified braneworld metric.  Also, the light
rays are gravitationally lensed by the central black hole as they escape the
disc, and we have shown here that such lensing can be different from the
Schwarzschild case for BBHs.   
In light of the results found here, it is not unreasonable to anticipate that
these effects could result in distinctive observational signatures for accretion discs.
This work is in progress.

\section*{Acknowledgements}
I would like to thank Ruth Gregory, Kris Beckwith, and Chris Done for helpful
comments and suggestions.  This work was supported by PPARC.

\end{document}